\documentclass[aps, prl,twocolumn,longbibliography,superscriptaddress,floatfix]{revtex4}
\usepackage[latin9]{inputenc}
\usepackage{amssymb}
\usepackage{graphicx}
\usepackage{amsmath}
\usepackage{color}
\usepackage{mathrsfs}
\usepackage{float}
\usepackage{indentfirst}
\usepackage{mathrsfs}
\usepackage{float}
\usepackage{indentfirst}
\usepackage{textcomp}
\usepackage{comment}
\usepackage{mathtools}
\usepackage{natbib,hyperref}
\usepackage{soul}

\begin{document}

\title{Effective bi-layer model Hamiltonian and density-matrix
renormalization group study for the high-Tc superconductivity in La$_{3}$Ni$%
_{2}$O$_{7}$ under high pressure}
\author{Yang Shen}
\affiliation{Key Laboratory of Artificial Structures and Quantum Control, School of
Physics and Astronomy, Shanghai Jiao Tong University, Shanghai 200240, China}
\author{Mingpu Qin}
\email{qinmingpu@sjtu.edu.cn}
\affiliation{Key Laboratory of Artificial Structures and Quantum Control, School of
Physics and Astronomy, Shanghai Jiao Tong University, Shanghai 200240, China}
\author{Guang-Ming Zhang}
\email{gmzhang@tsinghua.edu.cn}
\affiliation{State Key Laboratory of Low-Dimensional Quantum Physics and Department of
Physics, Tsinghua University, Beijing 100084, China}
\affiliation{Frontier Science Center for Quantum Information, Beijing 100084, China}
\date{\today}

\begin{abstract}
High-Tc superconductivity with possible $T_{c}\approx 80K$ has been reported
in the single crystal of $\text{La}_{3}\text{Ni}_{2}\text{O}_{7}$ under high
pressure. Based on the electronic structure given from the density
functional theory calculations, we propose an effective bi-layer model
Hamiltonian including both $3d_{z^{2}}$ and $3d_{x^{2}-y^{2}}$ orbital
electrons of the nickel cations. The main feature of the model is that the $%
3d_{z^{2}}$ electrons form inter-layer $\sigma$-bonding and anti-bonding
bands via the apical oxygen anions between the two layers, while the $%
3d_{x^{2}-y^{2}}$ electrons hybridize with the $3d_{z^{2}}$ electrons within
each NiO$_2$ plane. The chemical potential difference of these two orbital
electrons ensures that the $3d_{z^{2}}$ orbitals are close to half-filling
and the $3d_{x^{2}-y^{2}}$ orbitals are near quarter-filling. The strong
on-site Hubbard repulsion of the $3d_{z^{2}}$ orbital electrons gives rise
to an effective inter-layer antiferromagnetic spin super-exchange $J$.
Applying pressure can self-dope holes on the $3d_{z^{2}}$ orbitals with the
same amount of electrons doped on the $3d_{x^{2}-y^{2}}$ orbitals. By
performing numerical density-matrix renormalization group calculations on a
minimum setup and focusing on the limit of large $J$ and small doping of $%
3d_{z^{2}}$ orbitals, we find the superconducting instability on both the $%
3d_{z^{2}}$ and $3d_{x^{2}-y^{2}}$ orbitals by calculating the equal-time
spin singlet pair-pair correlation function. Our numerical results have
provided useful insights in the high-Tc superconductivity in single crystal
La$_3$Ni$_2$O$_7$ under high pressure.
\end{abstract}

\maketitle

\textbf{Introduction}. -The successfully synthesized infinite-layer
nickelate superconductors \cite%
{li2019superconductivity,nomura2022superconductivity,PhysRevLett.125.027001,PhysRevLett.125.147003,PhysRevMaterials.4.121801,https://doi.org/10.1002/adma.202104083,doi:10.1126/sciadv.abl9927}
have provided another platform for the study of the microscopic origin of
unconventional superconductivity. Similar to high-Tc cuprates, the
infinite-layer nickelates with nominal $Ni^{+}$ ($3d^{9}$) cations contain $%
3d_{x^{2}-y^{2}}$ orbital degrees of freedom on a quasi-two-dimensional Ni-O
square lattice\cite{anisimov1999electronic,lee2004infinite}. However, the
electronic states of oxygen $2p$ orbitals are far below the Fermi level and
have a much reduced $3d-2p$ mixing due to the larger separation of their
on-site energies, and their low-energy electronic structures are more likely
to fall into the Mott-Hubbard than the charge-transfer regime with a
super-exchange energy at least an order of magnitude smaller than in cuprates%
\cite{Jiang-Sawatzky}. As a consequence, the parent infinite-layer
nickelates may be modelled as a self-doped Mott insulator with two types of
charge carriers, and the low-temperature upturn of the electric resistivity
\cite{li2019superconductivity,Ikeda2016} arises from the magnetic spin
scattering between low-density conduction electrons from the rare earths and
localized Ni-3$d_{x^{2}-y^{2}}$ magnetic moments \cite%
{Zhang-Yang-Zhang2020,Wan-Zhang-Yang-Zhang2020}. Upon Sr doping, the
superconducting $T_{c}$ is just around 9-15 K in the Nd$_{0.8}$Sr$_{0.2}$NiO$%
_{2}$ thin films \cite{li2019superconductivity}, and the maximum $T_{c}$ of
31 K has been achieved in Pr$_{0.82}$Sr$_{0.18}$NiO$_{2}$ films under high
pressure \cite{jinguang}. So far no bulk crystals can be synthesized and
show superconductivity.

Recently, it has been reported that the superconductivity with possible $%
T_{c}\approx 80K$ is observed in the single crystal of $\text{La}_{3}\text{Ni%
}_{2}\text{O}_{7}$ with pressure between $14.0$ and $43.5$ GPa using
high-pressure resistance and mutual inductive magnetic susceptibility
measurements\cite{2023arXiv230509586S}. Density functional theory (DFT)
calculations for the two nearest intra-layer Ni cations in a bilayer
Ruddlesden-Popper (RP) phase\cite{Pardo-Warren2011,2023arXiv230509586S} have
suggested that both the $3d_{x^{2}-y^{2}}$ and $3d_{z^{2}}$ orbitals of Ni
cations strongly mix with oxygen $2p$ orbitals. The $3d_{z^{2}}$ orbitals
via the apical oxygen usually have a large inter-layer coupling due to the
quantum confinement of the NiO$_{2}$ bilayer in the structure, and the
resulting energy splitting of Ni cations can dramatically change the
distribution of the averaged valence state of $Ni^{2.5+}$. The numerical
results further indicated the superconductivity emerges coincidently with
the metallization of the $\sigma $-bonding bands under the Fermi level,
consisting of the $3d_{z^{2}}$ orbitals with the apical oxygen connecting
Ni-O bilayers \cite{PhysRevB.91.045132,tao1}. These distinct features are
important clues for the high-T$_{c}$ superconductivity in this
Ruddlesden-Popper double-layered perovskite nickelates, which are different
from the infinite-layer nickelate superconductors.

In this work, based on the DFT electronic structure \cite%
{Pardo-Warren2011,2023arXiv230509586S}, we propose an effective bi-layer
model Hamiltonian including both $3d_{z^{2}}$ and $3d_{x^{2}-y^{2}}$ orbital
electrons of the nickel cations, which are different from the single orbital
bi-layer Hubbard model \cite%
{Dagotto-1992,Bulut-1992,Kuroki2002,Maier-2011,Mishra2016,Kuroki2017,Maier-2019,Karakuzu-2021,Kato-2020}%
. The main feature of the model is that the $3d_{z^{2}}$ electrons form
inter-layer $\sigma $-bonding and anti-bonding bands via the apical oxygen
anions between two layers, while the $3d_{x^{2}-y^{2}}$ electrons hybridize
with the $3d_{z^{2}} $ electrons within each NiO$_{2}$ plane. Due to their
special spatial symmetries of two $e_{g}$ orbitals, the intra-layer hopping
of the $3d_{z^{2}}$ orbital electrons and the inter-layer hopping of the $%
3d_{x^{2}-y^{2}}$ orbital electrons are very small, and can be neglected.
The chemical potential difference of these two orbital electrons ensures
that the $3d_{z^{2}}$ orbitals are close to half-filling and the $%
3d_{x^{2}-y^{2}}$ orbitals are near quarter-filling. The strong on-site
Hubbard repulsion gives rise to an inter-layer antiferromagnetic
super-exchange of the $3d_{z^{2}}$ orbital electrons $J$. Applying pressure
can increase the coupling strength $J$ and self-dope additional holes on the
$3d_{z^{2}}$ orbitals with the same amount of electrons doped on the $%
3d_{x^{2}-y^{2}}$ orbitals. By performing numerical density-matrix
renormalization group (DMRG) calculations on a minimum one-dimensional setup
and focusing on the large $J$ and small doping of $3d_{z^{2}}$ orbital
limit, we observe a charge-density wave (CDW) for both $3d_{z^{2}}$ and $%
3d_{x^{2}-y^{2}}$ electrons but with different wave-lengths. We also find
instability of superconductivity for both orbitals from the equal-time spin
singlet pair-pair correlations. We attribute the pairing on $3d_{z^{2}}$
orbital to the formation of inter-layer singlet pairs, and the pairing on $%
3d_{x^{2}-y^{2}}$ orbitals from the hybridization of the two orbitals.

\begin{figure}[tbp]
\includegraphics[width=0.4\textwidth]{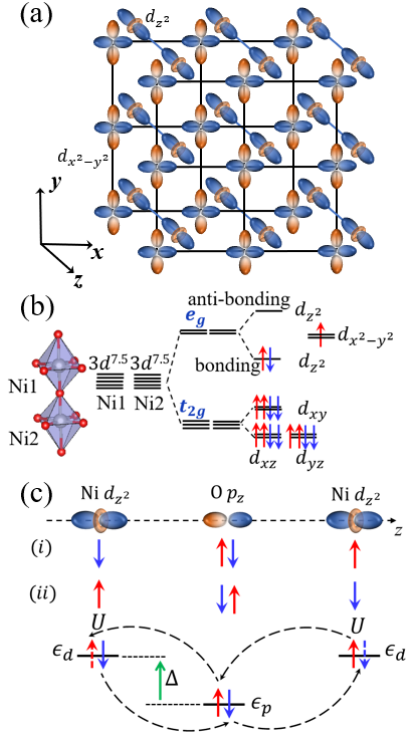}
\caption{(a) Schematic illustration of the $3d_{x^2-y^2}$ and $3d_{z^2}$
orbitals of Ni cations. We have omitted the $p_x$ and $p_y$ orbitals of
oxygen anions in the $xy$ plane and the $p_z$ orbitals of the apical oxygen
anions between the two layers. (b) The energy levels for two $3d$ orbitals
of Ni cations in one unit cell. (c) The antiferromagnetic spin
super-exchange coupling is resulted from the effective interactions between
the two inter-layer $3d_{z^2}$ orbitals of Ni cations via the apical oxygen $%
p_z$ orbitals. We just listed the two low-energy states and the higher
energy intermediate processes, leading to the effective interaction Eq.~(%
\protect\ref{AF_sup}).}
\label{Ni-structure}
\end{figure}

\textbf{Effective model Hamiltonian}. -Let us focus on the bi-layer RP bulk
single crystals of La$_{3}$Ni$_{2}$O$_{7}$ (see Fig.~\ref{Ni-structure} (a)
and (b)). A simple electron count gives a Ni$^{2.5+}$, i.e., 3d$^{7.5}$
state for both Ni cations, and the previous experiments indicated that La$%
_{3}$Ni$_{2}$O$_{7}$ is a paramagnetic metal. Ni$^{2.5+}$ is usually
believed to be given by mixed valence states of Ni$^{2+}$ (3d$^{8}$) and Ni$%
^{3+}$ (3d$^{7}$), corresponding to the half-filled of both $3d_{z^{2}}$ and
$3d_{x^{2}-y^{2}}$ orbitals and singly occupied $3d_{z^{2}}$ with empty $%
3d_{x^{2}-y^{2}}$ orbitals, respectively. With a bilayer RP phase, two $%
3d_{z^{2}}$ orbitals via apical oxygen anions usually have a large
inter-layer coupling due to the quantum confinement of the NiO$_{2}$ bilayer
in the structure, which causes a large energy splitting of Ni cations, see
Fig.~\ref{Ni-structure} (c). Then the distribution of the averaged valence
state of Ni$^{2.5+}$ can dramatically be changed.

We first consider an isolated Ni-O-Ni three-site system, which includes only
the half-filled $3d_{z^{2}}$ orbitals (see Fig.~\ref{Ni-structure} (c))
\begin{eqnarray}
H &=&v\sum_{\sigma }(d_{2\sigma }^{\dagger }p_{\sigma }-d_{1\sigma
}^{\dagger }p_{\sigma }+h.c)+\epsilon _{p}n_{p}  \notag \\
&&+\epsilon _{d}(n_{d1}+n_{d2})+Un_{d1\uparrow }n_{d1\downarrow
}+Un_{d2\uparrow }n_{d2\downarrow },  \label{ham_3site}
\end{eqnarray}%
where $v$ is the hybridization between the $3d_{z^{2}}$ orbital of nickel
and the $p_{z}$ orbital of apical oxygen, $\epsilon _{d}$ ($\epsilon _{p}$)
is the energy for the $3d_{z^{2}}$ ($p_{z}$) orbital, and $U$ is the on-site
Coulomb repulsion for the $3d_{z^{2}}$ orbital of nickel. Notice that the
signs for hybridization are different for the two $3d_{z^{2}}$ orbitals.
When $U\gg v,\epsilon _{p},\epsilon _{d}$, the lower energy configurations
are shown in (i) and (ii) in Fig.~\ref{Ni-structure} (c), where the oxygen $%
p_{z}$ orbital is doubly occupied and the $3d_{z^{2}}$ orbital is
half-filled. The state of these two $3d_{z^{2}}$ orbitals occupied by
parallel electrons has a higher energy. By considering the virtual
transition to higher energy states, we can derive an anti-ferromagnetic
super-exchanges between these two $3d_{z^{2}}$ orbitals as
\begin{equation}
H_{\text{eff}}=J\mathbf{S}_{1}\cdot \mathbf{S}_{2},  \label{AF_sup}
\end{equation}%
with
\begin{equation}
J=\frac{4v^{4}}{(U+\Delta )^{2}}(\frac{1}{U}+\frac{1}{U+\Delta }),\text{ }%
\Delta =\epsilon _{d}-\epsilon _{p}.
\end{equation}

Then a simplified bi-layer model Hamiltonian of a square planar coordinated
Ni cations consisting of the $3d_{z^{2}}$ and $3d_{x^{2}-y^{2}}$ orbitals
characterizes the effective low-energy physics,
\begin{eqnarray}
H &=&-t_{x^{2}-y^{2}}\sum_{\langle ij\rangle ,\sigma ,a=1,2}(c_{a,i,\sigma
}^{\dagger }c_{a,j,\sigma }+h.c)  \notag \\
&&-\mu _{x^{2}-y^{2}}\sum_{i,a=1,2}n_{a,i}^{c}+U\sum_{i,a=1,2}n_{a,i\uparrow
}^{c}n_{a,i\downarrow }^{c}  \notag \\
&&-t_{x^{2}-y^{2},z^{2}}\sum_{i,\sigma ,a=1,2}(d_{a,i,\sigma }^{\dagger }%
\widetilde{c}_{a,i,\sigma }+h.c)  \notag \\
&&-t_{z^{2}}\sum_{i,\sigma }(d_{1,i,\sigma }^{\dagger }d_{2,i,\sigma }+h.c)
\notag \\
&+&J\sum_{i}\mathbf{S}_{1,i}^{{d}}\cdot \mathbf{S}_{2,i}^{{d}}-\mu
_{z^{2}}\sum_{i,a}n_{a,i}^{d}  \label{ham2}
\end{eqnarray}%
with $c_{a,i,\sigma }^{\dagger }$ ($d_{a,i,\sigma }^{\dagger }$) creates a $%
3d_{x^{2}-y^{2}}$ ($3d_{z^{2}}$) electron at the $i$-th site for the layer $%
a=1,2$, $\widetilde{c}_{a,i,\sigma }=\left( c_{a,i+x,\sigma
}+c_{a,i-x,\sigma }-c_{a,i+y,\sigma }-c_{a,i-y,\sigma }\right) /2$, $%
t_{z^{2}}$ is the hopping for the $3d_{z^{2}}$ electron between the two
layers and is set as the energy unit, $t_{x^{2}-y^{2}}$ is the hopping for
the $3d_{x^{2}-y^{2}}$ electron in each layer, $t_{x^{2}-y^{2},z^{2}}$ is
the intra-layer hybridization of the $3d_{x^{2}-y^{2}}$ electron and $%
3d_{z^{2}}$ electrons between the nearest neighbors and its sign is opposite
for the vertical and horizontal directions\cite{2023arXiv230509586S}, and $%
\mu _{x^{2}-y^{2}}$ and $\mu _{z^{2}}$ are the chemical potentials for the $%
3d_{x^{2}-y^{2}}$ and $3d_{z^{2}}$ orbitals of nickels, respectively. Here $%
\mu _{z^{2}}$ should be much smaller than $\mu _{x^{2}-y^{2}}$ to ensure the
$3d_{z^{2}}$ orbital is near half-filling, while the $3d_{x^{2}-y^{2}}$
orbital is close to quarter-filling. In this model, we have ignored the
intra-layer (inter-layer) hopping for the $3d_{z^{2}}$ ($3d_{x^{2}-y^{2}}$)
orbitals and double occupancy of the $3d_{z^{2}}$ orbital is not allowed,
i.e., the local constraint $n_{a,i}^{d}=n_{a,i,\uparrow
}^{d}+n_{a,i,\downarrow }^{d}<2$ has been imposed. Actually this effective
bi-layer model Hamiltonian is different from the single orbital bi-layer
Hubbard model \cite%
{Dagotto-1992,Bulut-1992,Kuroki2002,Maier-2011,Mishra2016,Kuroki2017,Maier-2019,Karakuzu-2021,Kato-2020}%
, and the distinct low-energy physics can be expected.

When the intra-layer hybridization between two orbitals $%
t_{x^{2}-y^{2},z^{2}}$ is absent, the $3d_{z^{2}}$ orbitals are decoupled
with the $3d_{x^{2}-y^{2}}$ orbitals. Because the intra-layer hopping of the
$3d_{z^{2}}$ orbitals can be neglected, the ground state of the two-site
half-filled $3d_{z^{2}}$ orbitals is an isolated singlet. The filling of $%
3d_{x^{2}-y^{2}}$ is close to $n^{c}=1/2$ per lattice site, so the ground
state of the interacting $3d_{x^{2}-y^{2}}$ electrons behave as a
paramagnetic metal. With a finite hybridization $t_{x^{2}-y^{2},z^{2}}$, the
$3d_{z^{2}}$ electrons can hop on the lattice, and the isolated pairs of the
$3d_{z^{2}}$ electrons can gain coherence and the system may display
superconductivity in the large inter-layer coupling limit. This picture
shares a similarity to the superconductivity theory of the metallization of $%
\sigma $-bonding band \cite{PhysRevB.91.045132,tao1}.

\begin{figure}[t]
\includegraphics[width=0.46\textwidth]{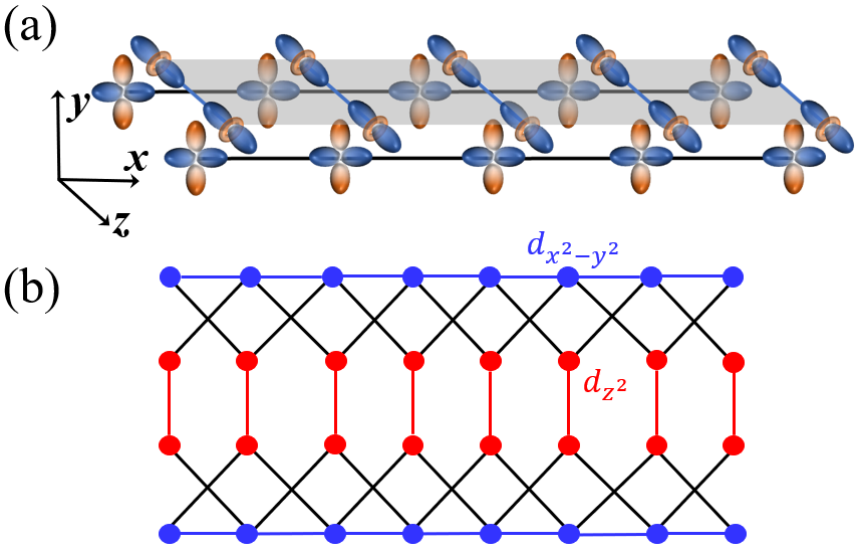}
\caption{(a) The minimum setup to capture the double-layer structure of La$%
_{3}$Ni$_{2}$O$_{7}$. (b) The lattice model used in the DMRG calculation.
Red and blue dots represent the 3$d_{z^2}$ and 3$d_{x^2-y^2}$ orbitals
respectively. We set the inter-layer hopping of 3$d_{z^2}$ orbital ($t_{z^2}$%
, red vertical lines) as the energy unit, and the hopping between 3$%
d_{x^2-y^2}$ orbitals $t_{x^{2}-y^{2}}=0.8$ (horizontal blue lines), the
hybridization between the 3$d_{z^2}$ and 3$d_{x^2-y^2}$ orbitals $%
t_{x^{2}-y^{2},z^{2}}=0.4$ (black cross lines). The supper exchange
interaction between the inter-layer 3$d_{z^2}$ orbitals is set as $J = 0.5$.}
\label{model_setup}
\end{figure}

\textbf{Numerical results of DMRG study}. -In order to explore the low
energy physics of the effective model Eq.~(\ref{ham2}), we employ the DMRG
method \cite{PhysRevLett.69.2863,PhysRevB.48.10345} to numerically solve a
minimum one-dimensional setup which captures the double-layer structure with
the length $L=32$ as shown in Fig.~\ref{model_setup}. According to the DFT
results\cite{Pardo-Warren2011,2023arXiv230509586S}, we choose the hopping
parameter $t_{z^{2}}$ as the energy unit, and $t_{x^{2}-y^{2}}=0.8$, $%
t_{x^{2}-y^{2},z^{2}}=0.4$. For the simplicity, we ignore the Hubbard
repulsion for the $3d_{x^{2}-y^{2}}$ orbitals because they are far away from
half-filled. Here we mainly focus on the large $J$ limit, so the local
antiferromagnetic spin coupling is set as $J=0.5$. Under the averaged
filling $n=3/4$, we have a scan of $\Delta _{\mu }$ = $\mu
_{x^{2}-y^{2}}-\mu _{z^{2}}$ to ensure that the $3d_{z^{2}}$ orbitals are $%
1/16$ hole doping away from half-filling and the $3d_{x^{2}-y^{2}}$ orbital
electrons have $9/16$ electron filling. With these chosen parameters, the $%
3d_{z^{2}}$ orbital electrons sit closely to the Mott-insulating limit,
while the $3d_{x^{2}-y^{2}}$ orbital electrons are in the large doping
limit, resembling the orbital selective Mott physics\cite%
{PhysRevLett.102.126401}. In our DMRG calculations, the maximum number of
the states is kept up to $m=18000$ with a truncation error $\epsilon
<5\times 10^{-6}$. In all the figures below, we will plot the numerical
results with large kept states $m$ to indicate the convergence of DMRG
calculations.

\begin{figure}[t]
\includegraphics[width=0.46\textwidth]{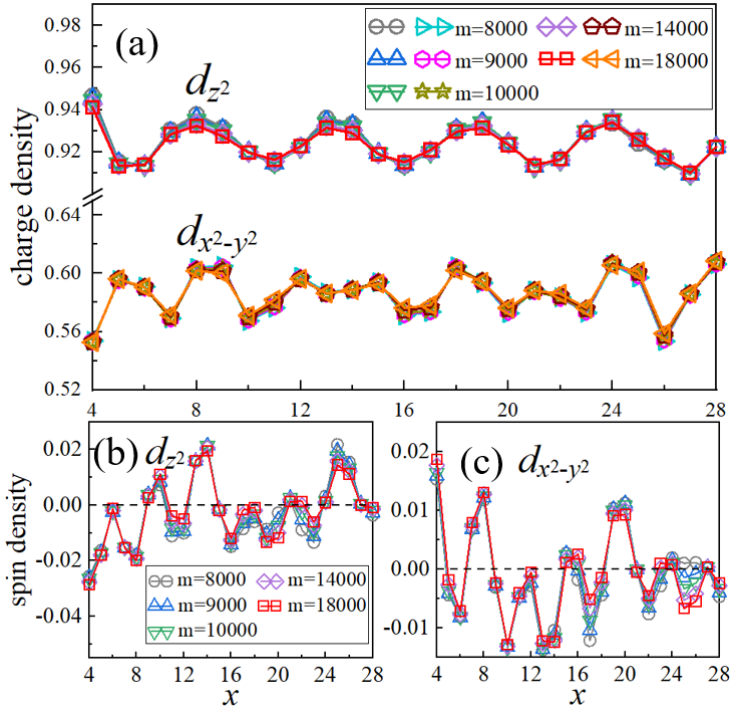}
\caption{(a) The charge density distribution of the $3d_{x^2-y^2}$ and $%
3d_{z^2}$ orbitals. The chemical potential difference $\Delta_{\protect\mu}$
is set $2.01$ to target the $1/16$ hole doping level on the $3d_{z^2}$
orbitals and $9/16$ electron filling on the $3d_{x^2-y^2}$ orbital. The
charge density wave pattern with wavelength $5$ can be seen in the $3d_{z^2}$
orbitals, while the pattern of the charge oscillation in the $3d_{x^2-y^2}$
orbitals has a wavelength $3$ roughly. (b) and (c) The short-ranged spin
density distribution of the $3d_{x^2-y^2}$ and $3d_{z^2}$ orbitals.}
\label{charge-spin}
\end{figure}

In Fig.~\ref{charge-spin} (a), we give the local charge density
distributions in real space of both the 3$d_{z^2}$ and 3$d_{x^2-y^2}$
electrons along one row, and the numerical results for other rows have the
same values. The obtained results with finite $m$ display nice convergence.
A CDW ordering with periodicity of $5$ is clearly seen for the 3$d_{z^2}$
orbital electrons, while there seems to be a CDW ordering with wavelength $3$
for the 3$d_{x^2-y^2}$ electrons, but the modulation is not as regular as
that in the 3$d_{z^2}$ orbitals.

In the DMRG method, the calculation of the spin-spin correlations is more
demanding. So we can also apply a pinning magnetic field with the strength $%
h_m = 0.5$ at one site in the left edge, which allows us to probe the
magnetic structure by calculating the local spin density\cite%
{PhysRevLett.99.127004}. In Fig.~\ref{charge-spin} (b) and (c), the local
spin densities of both the $3d_{x^2-y^2}$ and $3d_{z^2}$ orbitals are
displayed. Here the numerical results are shown for one row and the absolute
values of the different rows are almost the same. We can see that the spin
density of both orbital electrons is short-range (disordered), and the
system exhibits a paramagnetic behavior.

\begin{figure}[t]
\includegraphics[width=0.46\textwidth]{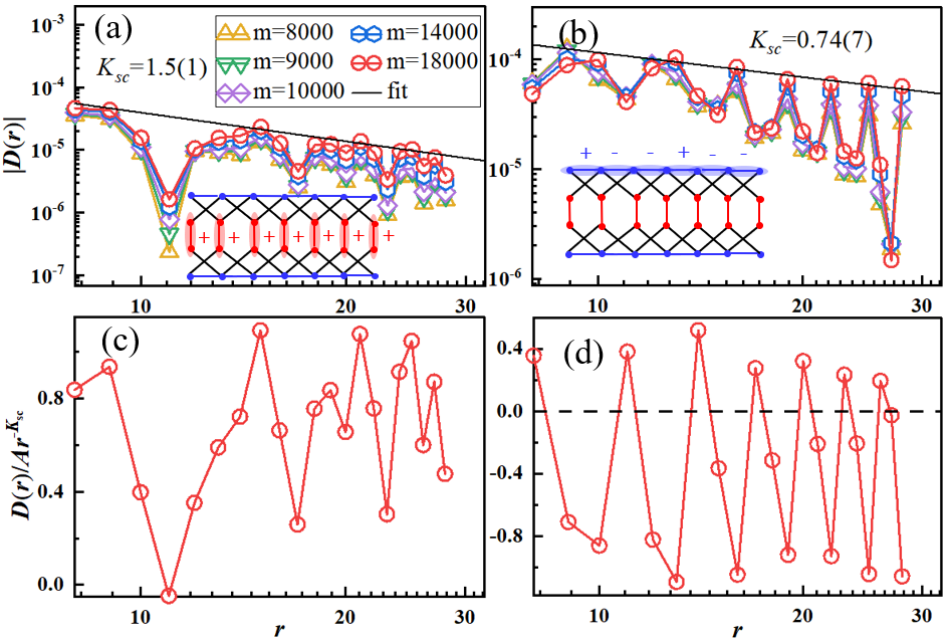}
\caption{The spin singlet pair-pair correlation functions for the minimum
setup model. The inter-layer $3d_{z^{2}}$ bond at $x = 4$ is set as the
reference bond in the calculation, and we can find oscillations with the
same period with the charge density. (a) The pair-pair correlation decay for
the inter-layer $3d_{z^{2}}$ bonds. The correlation is always positive, and
the envelope fitting gives an algebraically decay with exponent $%
K_{sc}=1.51(1)$. (b) The pair-pair correlation decay for the $%
3d_{x^{2}-y^{2}}$ bonds has a periodic sign structure of $+--$. The envelope
fitting gives an algebraically decay with exponent $K_{sc} = 0.74(7)$. In
(c) and (d), we display the pair-pair correlations divided by the fitted
envelope functions in (a) and (b) to clearly show the sign and oscillation
structure.}
\label{pair}
\end{figure}

In order to consider the superconductivity instability, we calculated the
equal-time spin singlet superconducting pair-pair correlation function
between bond $i$ (formed by site $(i,1)$ and $(i,2)$) and bond $j$ (formed
by site $(j,1)$ and $(j,2)$), which is defined as $D(i,j)=\langle \hat{\Delta%
}_{i}^{\dagger }\hat{\Delta}_{j}\rangle $, where
\begin{equation}
\hat{\Delta}_{i}^{\dagger }= (\hat{c}_{(i,1),\uparrow }^{\dagger }\hat{c}%
_{(i,2),\downarrow }^{\dagger }-\hat{c}_{(i,1),\downarrow }^{\dagger }\hat{c}%
_{(i,2),\uparrow }^{\dagger })/\sqrt{2}.
\end{equation}
We set the vertical 3$d_{z^{2}}$ bond as the reference bond in the
calculation of the pair-pair correlations. In Fig.~\ref{pair} (a), we show
the numerical results for the 3$d_{z^{2}}$ bonds, and find that the envelope
of the pair-pair correlation displays an algebraically decay $|D(r)|\sim
r^{-K_{sc}}$ with exponent $K_{sc} = 1.51(1)$. This pair-pair correlation is
always positive and oscillated in real space, consistent with the strong
local singlet pairings of the 3$d_{z^{2}}$ electrons. On the other hand, for
the 3$d_{x^{2}-y^{2}}$ electrons, the envelope of this pair-pair correlation
exhibits an algebraic decay with exponent $K_{sc} = 0.74(7)$ displayed in
Fig.~\ref{pair} (b). Moreover, this pair-pair correlation oscillates on the
lattice with a special sign structure $+--$.

Since both two fitted exponents $K_{sc}$ of the envelopes for 3$d_{z^{2}}$
and 3$d_{x^{2}-y^{2}}$ orbital electrons are smaller than $2$, the SC
correlations are strong enough so that a quasi-long-range order emerges,
implying the divergence of the static pair-pair SC susceptibility
characterized as
\begin{equation}
\chi _{sc}(T)\sim T^{-(2-K_{sc})},\text{ when }T\rightarrow 0.
\end{equation}
Moreover, the pair-pair correlations for $3d_{x^{2}-y^{2}}$ orbital
electrons exhibit oscillation with sign changes, so the emerged
superconductivity might be regarded as a pair-density wave ordered state. We
notice that the intra-layer pairing for the $3d_{x^{2}-y^{2}}$ orbitals is
stronger than the inter-layer pairing for the $3d_{z^{2}}$ orbitals. Though
the inter-layer super-exchange favors the formation of singlets for the $%
3d_{z^{2}}$ orbital, these isolated singlets need to hybridize with the
itinerant $3d_{x^{2}-y^{2}}$ orbital electrons to gain coherence. So it is
reasonable that the intra-layer pairing is stronger than the inter-layer
pairing, even though the latter is argued to be the origin of electron
pairing in the system.

\textbf{Conclusion}. -We have proposed an effective bi-layer model
Hamiltonian to describe the low energy physics of high-Tc superconductivity $%
\text{La}_{3}\text{Ni}_{2}\text{O}_{7}$ under high pressure \cite%
{2023arXiv230509586S}. In this effective model, we have argued that the $%
\sigma $-bonding band formed from the $3d_{z^{2}}$ orbitals via the apical
oxygen can be metallized due to the hybridization with the itinerant $%
3d_{x^{2}-y^{2}}$ orbitals, displaying unconventional high-Tc
superconductivity. We also performed DMRG study on a minimum one-dimensional
setup with the length $L=32$. The DMRG results have shown instability of
CDW-modulated superconductivity. The dominant spin singlet pair-pair
correlation is from the $3d_{x^{2}-y^{2}}$ orbitals, displaying a
pair-density-wave quasi-long-range order. Though this numerical calculation
for a minimum setup, we would like to argue that the obtained properties can
be used to justify the validity of the proposed effective bi-layer model
Hamiltonian for the understanding the high-Tc superconductivity in $\text{La}%
_{3}\text{Ni}_{2}\text{O}_{7}$ under high pressure.

\textbf{Acknowledgments}. - G.M. Zhang is grateful to Meng Wang and Fu-Chun
Zhang for their useful discussions. Y. Shen and M.P. Qin thank Weidong Luo
for his generosity to provide computational resources for this work. G.M.
Zhang acknowledges the support from the National Key Research and
Development Program of MOST of China (2017YFA0302902). M. P. Qin
acknowledges the support from the National Key Research and Development
Program of MOST of China (2022YFA1405400), the National Natural Science
Foundation of China (Grant No. 12274290) and the sponsorship from Yangyang
Development Fund. All the DMRG calculations are carried out with iTensor
library \cite{10.21468/SciPostPhysCodeb.4}.

\textbf{Note added}. During the preparation of this work, several
theoretical studies \cite%
{Yao,QiangHuaWang,Dagotto,Lechermann,Kuroki,JiangPingHu} appeared on arXiv
and the electronic structures and possible pairing instabilities of the
high-Tc superconductivity of La$_3$Ni$_2$O$_7$ under high pressure are
independently discussed.

\bibliography{Ni-model}

\begin{thebibliography}{38}
\expandafter\ifx\csname natexlab\endcsname\relax\def\natexlab#1{#1}\fi
\expandafter\ifx\csname bibnamefont\endcsname\relax
  \def\bibnamefont#1{#1}\fi
\expandafter\ifx\csname bibfnamefont\endcsname\relax
  \def\bibfnamefont#1{#1}\fi
\expandafter\ifx\csname citenamefont\endcsname\relax
  \def\citenamefont#1{#1}\fi
\expandafter\ifx\csname url\endcsname\relax
  \def\url#1{\texttt{#1}}\fi
\expandafter\ifx\csname urlprefix\endcsname\relax\def\urlprefix{URL }\fi
\providecommand{\bibinfo}[2]{#2}
\providecommand{\eprint}[2][]{\url{#2}}

\bibitem[{\citenamefont{Li et~al.}(2019)\citenamefont{Li, Lee, Wang, Osada,
  Crossley, Lee, Cui, Hikita, and Hwang}}]{li2019superconductivity}
\bibinfo{author}{\bibfnamefont{D.}~\bibnamefont{Li}},
  \bibinfo{author}{\bibfnamefont{K.}~\bibnamefont{Lee}},
  \bibinfo{author}{\bibfnamefont{B.~Y.} \bibnamefont{Wang}},
  \bibinfo{author}{\bibfnamefont{M.}~\bibnamefont{Osada}},
  \bibinfo{author}{\bibfnamefont{S.}~\bibnamefont{Crossley}},
  \bibinfo{author}{\bibfnamefont{H.~R.} \bibnamefont{Lee}},
  \bibinfo{author}{\bibfnamefont{Y.}~\bibnamefont{Cui}},
  \bibinfo{author}{\bibfnamefont{Y.}~\bibnamefont{Hikita}}, \bibnamefont{and}
  \bibinfo{author}{\bibfnamefont{H.~Y.} \bibnamefont{Hwang}},
  \bibinfo{journal}{Nature} \textbf{\bibinfo{volume}{572}},
  \bibinfo{pages}{624} (\bibinfo{year}{2019}).

\bibitem[{\citenamefont{Nomura and Arita}(2022)}]{nomura2022superconductivity}
\bibinfo{author}{\bibfnamefont{Y.}~\bibnamefont{Nomura}} \bibnamefont{and}
  \bibinfo{author}{\bibfnamefont{R.}~\bibnamefont{Arita}},
  \bibinfo{journal}{Reports on Progress in Physics}
  \textbf{\bibinfo{volume}{85}}, \bibinfo{pages}{052501}
  (\bibinfo{year}{2022}),
  \urlprefix\url{https://dx.doi.org/10.1088/1361-6633/ac5a60}.

\bibitem[{\citenamefont{Li et~al.}(2020)\citenamefont{Li, Wang, Lee, Harvey,
  Osada, Goodge, Kourkoutis, and Hwang}}]{PhysRevLett.125.027001}
\bibinfo{author}{\bibfnamefont{D.}~\bibnamefont{Li}},
  \bibinfo{author}{\bibfnamefont{B.~Y.} \bibnamefont{Wang}},
  \bibinfo{author}{\bibfnamefont{K.}~\bibnamefont{Lee}},
  \bibinfo{author}{\bibfnamefont{S.~P.} \bibnamefont{Harvey}},
  \bibinfo{author}{\bibfnamefont{M.}~\bibnamefont{Osada}},
  \bibinfo{author}{\bibfnamefont{B.~H.} \bibnamefont{Goodge}},
  \bibinfo{author}{\bibfnamefont{L.~F.} \bibnamefont{Kourkoutis}},
  \bibnamefont{and} \bibinfo{author}{\bibfnamefont{H.~Y.} \bibnamefont{Hwang}},
  \bibinfo{journal}{Phys. Rev. Lett.} \textbf{\bibinfo{volume}{125}},
  \bibinfo{pages}{027001} (\bibinfo{year}{2020}),
  \urlprefix\url{https://link.aps.org/doi/10.1103/PhysRevLett.125.027001}.

\bibitem[{\citenamefont{Zeng et~al.}(2020)\citenamefont{Zeng, Tang, Yin, Li,
  Li, Huang, Hu, Liu, Omar, Jani et~al.}}]{PhysRevLett.125.147003}
\bibinfo{author}{\bibfnamefont{S.}~\bibnamefont{Zeng}},
  \bibinfo{author}{\bibfnamefont{C.~S.} \bibnamefont{Tang}},
  \bibinfo{author}{\bibfnamefont{X.}~\bibnamefont{Yin}},
  \bibinfo{author}{\bibfnamefont{C.}~\bibnamefont{Li}},
  \bibinfo{author}{\bibfnamefont{M.}~\bibnamefont{Li}},
  \bibinfo{author}{\bibfnamefont{Z.}~\bibnamefont{Huang}},
  \bibinfo{author}{\bibfnamefont{J.}~\bibnamefont{Hu}},
  \bibinfo{author}{\bibfnamefont{W.}~\bibnamefont{Liu}},
  \bibinfo{author}{\bibfnamefont{G.~J.} \bibnamefont{Omar}},
  \bibinfo{author}{\bibfnamefont{H.}~\bibnamefont{Jani}}, \bibnamefont{et~al.},
  \bibinfo{journal}{Phys. Rev. Lett.} \textbf{\bibinfo{volume}{125}},
  \bibinfo{pages}{147003} (\bibinfo{year}{2020}),
  \urlprefix\url{https://link.aps.org/doi/10.1103/PhysRevLett.125.147003}.

\bibitem[{\citenamefont{Osada et~al.}(2020)\citenamefont{Osada, Wang, Lee, Li,
  and Hwang}}]{PhysRevMaterials.4.121801}
\bibinfo{author}{\bibfnamefont{M.}~\bibnamefont{Osada}},
  \bibinfo{author}{\bibfnamefont{B.~Y.} \bibnamefont{Wang}},
  \bibinfo{author}{\bibfnamefont{K.}~\bibnamefont{Lee}},
  \bibinfo{author}{\bibfnamefont{D.}~\bibnamefont{Li}}, \bibnamefont{and}
  \bibinfo{author}{\bibfnamefont{H.~Y.} \bibnamefont{Hwang}},
  \bibinfo{journal}{Phys. Rev. Materials} \textbf{\bibinfo{volume}{4}},
  \bibinfo{pages}{121801} (\bibinfo{year}{2020}),
  \urlprefix\url{https://link.aps.org/doi/10.1103/PhysRevMaterials.4.121801}.

\bibitem[{\citenamefont{Osada et~al.}(2021)\citenamefont{Osada, Wang, Goodge,
  Harvey, Lee, Li, Kourkoutis, and
  Hwang}}]{https://doi.org/10.1002/adma.202104083}
\bibinfo{author}{\bibfnamefont{M.}~\bibnamefont{Osada}},
  \bibinfo{author}{\bibfnamefont{B.~Y.} \bibnamefont{Wang}},
  \bibinfo{author}{\bibfnamefont{B.~H.} \bibnamefont{Goodge}},
  \bibinfo{author}{\bibfnamefont{S.~P.} \bibnamefont{Harvey}},
  \bibinfo{author}{\bibfnamefont{K.}~\bibnamefont{Lee}},
  \bibinfo{author}{\bibfnamefont{D.}~\bibnamefont{Li}},
  \bibinfo{author}{\bibfnamefont{L.~F.} \bibnamefont{Kourkoutis}},
  \bibnamefont{and} \bibinfo{author}{\bibfnamefont{H.~Y.} \bibnamefont{Hwang}},
  \bibinfo{journal}{Advanced Materials} \textbf{\bibinfo{volume}{33}},
  \bibinfo{pages}{2104083} (\bibinfo{year}{2021}),
  \urlprefix\url{https://onlinelibrary.wiley.com/doi/abs/10.1002/adma.202104083}.

\bibitem[{\citenamefont{Zeng et~al.}(2022)\citenamefont{Zeng, Li, Chow, Cao,
  Zhang, Tang, Yin, Lim, Hu, Yang et~al.}}]{doi:10.1126/sciadv.abl9927}
\bibinfo{author}{\bibfnamefont{S.}~\bibnamefont{Zeng}},
  \bibinfo{author}{\bibfnamefont{C.}~\bibnamefont{Li}},
  \bibinfo{author}{\bibfnamefont{L.~E.} \bibnamefont{Chow}},
  \bibinfo{author}{\bibfnamefont{Y.}~\bibnamefont{Cao}},
  \bibinfo{author}{\bibfnamefont{Z.}~\bibnamefont{Zhang}},
  \bibinfo{author}{\bibfnamefont{C.~S.} \bibnamefont{Tang}},
  \bibinfo{author}{\bibfnamefont{X.}~\bibnamefont{Yin}},
  \bibinfo{author}{\bibfnamefont{Z.~S.} \bibnamefont{Lim}},
  \bibinfo{author}{\bibfnamefont{J.}~\bibnamefont{Hu}},
  \bibinfo{author}{\bibfnamefont{P.}~\bibnamefont{Yang}}, \bibnamefont{et~al.},
  \bibinfo{journal}{Science Advances} \textbf{\bibinfo{volume}{8}},
  \bibinfo{pages}{eabl9927} (\bibinfo{year}{2022}),
  \eprint{https://www.science.org/doi/pdf/10.1126/sciadv.abl9927},
  \urlprefix\url{https://www.science.org/doi/abs/10.1126/sciadv.abl9927}.

\bibitem[{\citenamefont{Anisimov et~al.}(1999)\citenamefont{Anisimov,
  Bukhvalov, and Rice}}]{anisimov1999electronic}
\bibinfo{author}{\bibfnamefont{V.~I.} \bibnamefont{Anisimov}},
  \bibinfo{author}{\bibfnamefont{D.}~\bibnamefont{Bukhvalov}},
  \bibnamefont{and} \bibinfo{author}{\bibfnamefont{T.~M.} \bibnamefont{Rice}},
  \bibinfo{journal}{Phys. Rev. B} \textbf{\bibinfo{volume}{59}},
  \bibinfo{pages}{7901} (\bibinfo{year}{1999}),
  \urlprefix\url{https://link.aps.org/doi/10.1103/PhysRevB.59.7901}.

\bibitem[{\citenamefont{Lee and Pickett}(2004)}]{lee2004infinite}
\bibinfo{author}{\bibfnamefont{K.-W.} \bibnamefont{Lee}} \bibnamefont{and}
  \bibinfo{author}{\bibfnamefont{W.~E.} \bibnamefont{Pickett}},
  \bibinfo{journal}{Phys. Rev. B} \textbf{\bibinfo{volume}{70}},
  \bibinfo{pages}{165109} (\bibinfo{year}{2004}),
  \urlprefix\url{https://link.aps.org/doi/10.1103/PhysRevB.70.165109}.

\bibitem[{\citenamefont{Jiang et~al.}(2020)\citenamefont{Jiang, Berciu, and
  Sawatzky}}]{Jiang-Sawatzky}
\bibinfo{author}{\bibfnamefont{M.}~\bibnamefont{Jiang}},
  \bibinfo{author}{\bibfnamefont{M.}~\bibnamefont{Berciu}}, \bibnamefont{and}
  \bibinfo{author}{\bibfnamefont{G.~A.} \bibnamefont{Sawatzky}},
  \bibinfo{journal}{Phys. Rev. Lett.} \textbf{\bibinfo{volume}{124}},
  \bibinfo{pages}{207004} (\bibinfo{year}{2020}),
  \urlprefix\url{https://link.aps.org/doi/10.1103/PhysRevLett.124.207004}.

\bibitem[{\citenamefont{Ikeda et~al.}(2016)\citenamefont{Ikeda, Krockenberger,
  Irie, Naito, and Yamamoto}}]{Ikeda2016}
\bibinfo{author}{\bibfnamefont{A.}~\bibnamefont{Ikeda}},
  \bibinfo{author}{\bibfnamefont{Y.}~\bibnamefont{Krockenberger}},
  \bibinfo{author}{\bibfnamefont{H.}~\bibnamefont{Irie}},
  \bibinfo{author}{\bibfnamefont{M.}~\bibnamefont{Naito}}, \bibnamefont{and}
  \bibinfo{author}{\bibfnamefont{H.}~\bibnamefont{Yamamoto}},
  \bibinfo{journal}{Applied Physics Express} \textbf{\bibinfo{volume}{9}},
  \bibinfo{pages}{061101} (\bibinfo{year}{2016}),
  \urlprefix\url{https://dx.doi.org/10.7567/APEX.9.061101}.

\bibitem[{\citenamefont{Zhang et~al.}(2020)\citenamefont{Zhang, Yang, and
  Zhang}}]{Zhang-Yang-Zhang2020}
\bibinfo{author}{\bibfnamefont{G.-M.} \bibnamefont{Zhang}},
  \bibinfo{author}{\bibfnamefont{Y.-F.} \bibnamefont{Yang}}, \bibnamefont{and}
  \bibinfo{author}{\bibfnamefont{F.-C.} \bibnamefont{Zhang}},
  \bibinfo{journal}{Phys. Rev. B} \textbf{\bibinfo{volume}{101}},
  \bibinfo{pages}{020501} (\bibinfo{year}{2020}),
  \urlprefix\url{https://link.aps.org/doi/10.1103/PhysRevB.101.020501}.

\bibitem[{\citenamefont{Wang et~al.}(2020)\citenamefont{Wang, Zhang, Yang, and
  Zhang}}]{Wan-Zhang-Yang-Zhang2020}
\bibinfo{author}{\bibfnamefont{Z.}~\bibnamefont{Wang}},
  \bibinfo{author}{\bibfnamefont{G.-M.} \bibnamefont{Zhang}},
  \bibinfo{author}{\bibfnamefont{Y.-f.} \bibnamefont{Yang}}, \bibnamefont{and}
  \bibinfo{author}{\bibfnamefont{F.-C.} \bibnamefont{Zhang}},
  \bibinfo{journal}{Phys. Rev. B} \textbf{\bibinfo{volume}{102}},
  \bibinfo{pages}{220501} (\bibinfo{year}{2020}),
  \urlprefix\url{https://link.aps.org/doi/10.1103/PhysRevB.102.220501}.

\bibitem[{\citenamefont{Wang et~al.}(2022)\citenamefont{Wang, Yang, Yang, Chen,
  Zhang, Zhang, Zhu, Uwatoko, Gu, Dong et~al.}}]{jinguang}
\bibinfo{author}{\bibfnamefont{N.~N.} \bibnamefont{Wang}},
  \bibinfo{author}{\bibfnamefont{M.~W.} \bibnamefont{Yang}},
  \bibinfo{author}{\bibfnamefont{Z.}~\bibnamefont{Yang}},
  \bibinfo{author}{\bibfnamefont{K.~Y.} \bibnamefont{Chen}},
  \bibinfo{author}{\bibfnamefont{H.}~\bibnamefont{Zhang}},
  \bibinfo{author}{\bibfnamefont{Q.~H.} \bibnamefont{Zhang}},
  \bibinfo{author}{\bibfnamefont{Z.~H.} \bibnamefont{Zhu}},
  \bibinfo{author}{\bibfnamefont{Y.}~\bibnamefont{Uwatoko}},
  \bibinfo{author}{\bibfnamefont{L.}~\bibnamefont{Gu}},
  \bibinfo{author}{\bibfnamefont{X.~L.} \bibnamefont{Dong}},
  \bibnamefont{et~al.}, \bibinfo{journal}{Nature Communications}
  \textbf{\bibinfo{volume}{13}} (\bibinfo{year}{2022}),
  \urlprefix\url{https://doi.org/10.1038%2Fs41467-022-32065-x}.

\bibitem[{\citenamefont{{Sun} et~al.}(2023)\citenamefont{{Sun}, {Huo}, {Hu},
  {Li}, {Han}, {Tang}, {Mao}, {Yang}, {Wang}, {Cheng}
  et~al.}}]{2023arXiv230509586S}
\bibinfo{author}{\bibfnamefont{H.}~\bibnamefont{{Sun}}},
  \bibinfo{author}{\bibfnamefont{M.}~\bibnamefont{{Huo}}},
  \bibinfo{author}{\bibfnamefont{X.}~\bibnamefont{{Hu}}},
  \bibinfo{author}{\bibfnamefont{J.}~\bibnamefont{{Li}}},
  \bibinfo{author}{\bibfnamefont{Y.}~\bibnamefont{{Han}}},
  \bibinfo{author}{\bibfnamefont{L.}~\bibnamefont{{Tang}}},
  \bibinfo{author}{\bibfnamefont{Z.}~\bibnamefont{{Mao}}},
  \bibinfo{author}{\bibfnamefont{P.}~\bibnamefont{{Yang}}},
  \bibinfo{author}{\bibfnamefont{B.}~\bibnamefont{{Wang}}},
  \bibinfo{author}{\bibfnamefont{J.}~\bibnamefont{{Cheng}}},
  \bibnamefont{et~al.}, \bibinfo{journal}{arXiv e-prints}
  \bibinfo{eid}{arXiv:2305.09586} (\bibinfo{year}{2023}), \eprint{2305.09586}.

\bibitem[{\citenamefont{Pardo and Pickett}(2011)}]{Pardo-Warren2011}
\bibinfo{author}{\bibfnamefont{V.}~\bibnamefont{Pardo}} \bibnamefont{and}
  \bibinfo{author}{\bibfnamefont{W.~E.} \bibnamefont{Pickett}},
  \bibinfo{journal}{Phys. Rev. B} \textbf{\bibinfo{volume}{83}},
  \bibinfo{pages}{245128} (\bibinfo{year}{2011}),
  \urlprefix\url{https://link.aps.org/doi/10.1103/PhysRevB.83.245128}.

\bibitem[{\citenamefont{Gao et~al.}(2015{\natexlab{a}})\citenamefont{Gao, Lu,
  and Xiang}}]{PhysRevB.91.045132}
\bibinfo{author}{\bibfnamefont{M.}~\bibnamefont{Gao}},
  \bibinfo{author}{\bibfnamefont{Z.-Y.} \bibnamefont{Lu}}, \bibnamefont{and}
  \bibinfo{author}{\bibfnamefont{T.}~\bibnamefont{Xiang}},
  \bibinfo{journal}{Phys. Rev. B} \textbf{\bibinfo{volume}{91}},
  \bibinfo{pages}{045132} (\bibinfo{year}{2015}{\natexlab{a}}),
  \urlprefix\url{https://link.aps.org/doi/10.1103/PhysRevB.91.045132}.

\bibitem[{\citenamefont{Gao et~al.}(2015{\natexlab{b}})\citenamefont{Gao, Lu,
  and Xiang}}]{tao1}
\bibinfo{author}{\bibfnamefont{M.}~\bibnamefont{Gao}},
  \bibinfo{author}{\bibfnamefont{Z.-Y.} \bibnamefont{Lu}}, \bibnamefont{and}
  \bibinfo{author}{\bibfnamefont{T.}~\bibnamefont{Xiang}},
  \bibinfo{journal}{Physics} \textbf{\bibinfo{volume}{44}},
  \bibinfo{pages}{421} (\bibinfo{year}{2015}{\natexlab{b}}),
  \urlprefix\url{https://wuli.iphy.ac.cn/article/doi/10.7693/wl20150701}.

\bibitem[{\citenamefont{Dagotto et~al.}(1992)\citenamefont{Dagotto, Riera, and
  Scalapino}}]{Dagotto-1992}
\bibinfo{author}{\bibfnamefont{E.}~\bibnamefont{Dagotto}},
  \bibinfo{author}{\bibfnamefont{J.}~\bibnamefont{Riera}}, \bibnamefont{and}
  \bibinfo{author}{\bibfnamefont{D.}~\bibnamefont{Scalapino}},
  \bibinfo{journal}{Phys. Rev. B} \textbf{\bibinfo{volume}{45}},
  \bibinfo{pages}{5744} (\bibinfo{year}{1992}),
  \urlprefix\url{https://link.aps.org/doi/10.1103/PhysRevB.45.5744}.

\bibitem[{\citenamefont{Bulut et~al.}(1992)\citenamefont{Bulut, Scalapino, and
  Scalettar}}]{Bulut-1992}
\bibinfo{author}{\bibfnamefont{N.}~\bibnamefont{Bulut}},
  \bibinfo{author}{\bibfnamefont{D.~J.} \bibnamefont{Scalapino}},
  \bibnamefont{and} \bibinfo{author}{\bibfnamefont{R.~T.}
  \bibnamefont{Scalettar}}, \bibinfo{journal}{Phys. Rev. B}
  \textbf{\bibinfo{volume}{45}}, \bibinfo{pages}{5577} (\bibinfo{year}{1992}),
  \urlprefix\url{https://link.aps.org/doi/10.1103/PhysRevB.45.5577}.

\bibitem[{\citenamefont{Kuroki et~al.}(2002)\citenamefont{Kuroki, Kimura, and
  Arita}}]{Kuroki2002}
\bibinfo{author}{\bibfnamefont{K.}~\bibnamefont{Kuroki}},
  \bibinfo{author}{\bibfnamefont{T.}~\bibnamefont{Kimura}}, \bibnamefont{and}
  \bibinfo{author}{\bibfnamefont{R.}~\bibnamefont{Arita}},
  \bibinfo{journal}{Phys. Rev. B} \textbf{\bibinfo{volume}{66}},
  \bibinfo{pages}{184508} (\bibinfo{year}{2002}),
  \urlprefix\url{https://link.aps.org/doi/10.1103/PhysRevB.66.184508}.

\bibitem[{\citenamefont{Maier and Scalapino}(2011)}]{Maier-2011}
\bibinfo{author}{\bibfnamefont{T.~A.} \bibnamefont{Maier}} \bibnamefont{and}
  \bibinfo{author}{\bibfnamefont{D.~J.} \bibnamefont{Scalapino}},
  \bibinfo{journal}{Phys. Rev. B} \textbf{\bibinfo{volume}{84}},
  \bibinfo{pages}{180513} (\bibinfo{year}{2011}),
  \urlprefix\url{https://link.aps.org/doi/10.1103/PhysRevB.84.180513}.

\bibitem[{\citenamefont{Mishra et~al.}(2016)\citenamefont{Mishra, Scalapino,
  and Maier}}]{Mishra2016}
\bibinfo{author}{\bibfnamefont{V.}~\bibnamefont{Mishra}},
  \bibinfo{author}{\bibfnamefont{D.~J.} \bibnamefont{Scalapino}},
  \bibnamefont{and} \bibinfo{author}{\bibfnamefont{T.~A.} \bibnamefont{Maier}},
  \bibinfo{journal}{Scientific Reports} \textbf{\bibinfo{volume}{6}}
  (\bibinfo{year}{2016}), \urlprefix\url{https://doi.org/10.1038%2Fsrep32078}.

\bibitem[{\citenamefont{Nakata et~al.}(2017)\citenamefont{Nakata, Ogura, Usui,
  and Kuroki}}]{Kuroki2017}
\bibinfo{author}{\bibfnamefont{M.}~\bibnamefont{Nakata}},
  \bibinfo{author}{\bibfnamefont{D.}~\bibnamefont{Ogura}},
  \bibinfo{author}{\bibfnamefont{H.}~\bibnamefont{Usui}}, \bibnamefont{and}
  \bibinfo{author}{\bibfnamefont{K.}~\bibnamefont{Kuroki}},
  \bibinfo{journal}{Phys. Rev. B} \textbf{\bibinfo{volume}{95}},
  \bibinfo{pages}{214509} (\bibinfo{year}{2017}),
  \urlprefix\url{https://link.aps.org/doi/10.1103/PhysRevB.95.214509}.

\bibitem[{\citenamefont{Maier et~al.}(2019)\citenamefont{Maier, Mishra,
  Balduzzi, and Scalapino}}]{Maier-2019}
\bibinfo{author}{\bibfnamefont{T.~A.} \bibnamefont{Maier}},
  \bibinfo{author}{\bibfnamefont{V.}~\bibnamefont{Mishra}},
  \bibinfo{author}{\bibfnamefont{G.}~\bibnamefont{Balduzzi}}, \bibnamefont{and}
  \bibinfo{author}{\bibfnamefont{D.~J.} \bibnamefont{Scalapino}},
  \bibinfo{journal}{Phys. Rev. B} \textbf{\bibinfo{volume}{99}},
  \bibinfo{pages}{140504} (\bibinfo{year}{2019}),
  \urlprefix\url{https://link.aps.org/doi/10.1103/PhysRevB.99.140504}.

\bibitem[{\citenamefont{Karakuzu et~al.}(2021)\citenamefont{Karakuzu, Johnston,
  and Maier}}]{Karakuzu-2021}
\bibinfo{author}{\bibfnamefont{S.}~\bibnamefont{Karakuzu}},
  \bibinfo{author}{\bibfnamefont{S.}~\bibnamefont{Johnston}}, \bibnamefont{and}
  \bibinfo{author}{\bibfnamefont{T.~A.} \bibnamefont{Maier}},
  \bibinfo{journal}{Phys. Rev. B} \textbf{\bibinfo{volume}{104}},
  \bibinfo{pages}{245109} (\bibinfo{year}{2021}),
  \urlprefix\url{https://link.aps.org/doi/10.1103/PhysRevB.104.245109}.

\bibitem[{\citenamefont{Kato and Kuroki}(2020)}]{Kato-2020}
\bibinfo{author}{\bibfnamefont{D.}~\bibnamefont{Kato}} \bibnamefont{and}
  \bibinfo{author}{\bibfnamefont{K.}~\bibnamefont{Kuroki}},
  \bibinfo{journal}{Phys. Rev. Res.} \textbf{\bibinfo{volume}{2}},
  \bibinfo{pages}{023156} (\bibinfo{year}{2020}),
  \urlprefix\url{https://link.aps.org/doi/10.1103/PhysRevResearch.2.023156}.

\bibitem[{\citenamefont{White}(1992)}]{PhysRevLett.69.2863}
\bibinfo{author}{\bibfnamefont{S.~R.} \bibnamefont{White}},
  \bibinfo{journal}{Phys. Rev. Lett.} \textbf{\bibinfo{volume}{69}},
  \bibinfo{pages}{2863} (\bibinfo{year}{1992}),
  \urlprefix\url{https://link.aps.org/doi/10.1103/PhysRevLett.69.2863}.

\bibitem[{\citenamefont{White}(1993)}]{PhysRevB.48.10345}
\bibinfo{author}{\bibfnamefont{S.~R.} \bibnamefont{White}},
  \bibinfo{journal}{Phys. Rev. B} \textbf{\bibinfo{volume}{48}},
  \bibinfo{pages}{10345} (\bibinfo{year}{1993}),
  \urlprefix\url{https://link.aps.org/doi/10.1103/PhysRevB.48.10345}.

\bibitem[{\citenamefont{de' Medici et~al.}(2009)\citenamefont{de' Medici,
  Hassan, Capone, and Dai}}]{PhysRevLett.102.126401}
\bibinfo{author}{\bibfnamefont{L.}~\bibnamefont{de' Medici}},
  \bibinfo{author}{\bibfnamefont{S.~R.} \bibnamefont{Hassan}},
  \bibinfo{author}{\bibfnamefont{M.}~\bibnamefont{Capone}}, \bibnamefont{and}
  \bibinfo{author}{\bibfnamefont{X.}~\bibnamefont{Dai}},
  \bibinfo{journal}{Phys. Rev. Lett.} \textbf{\bibinfo{volume}{102}},
  \bibinfo{pages}{126401} (\bibinfo{year}{2009}),
  \urlprefix\url{https://link.aps.org/doi/10.1103/PhysRevLett.102.126401}.

\bibitem[{\citenamefont{White and Chernyshev}(2007)}]{PhysRevLett.99.127004}
\bibinfo{author}{\bibfnamefont{S.~R.} \bibnamefont{White}} \bibnamefont{and}
  \bibinfo{author}{\bibfnamefont{A.~L.} \bibnamefont{Chernyshev}},
  \bibinfo{journal}{Phys. Rev. Lett.} \textbf{\bibinfo{volume}{99}},
  \bibinfo{pages}{127004} (\bibinfo{year}{2007}),
  \urlprefix\url{https://link.aps.org/doi/10.1103/PhysRevLett.99.127004}.

\bibitem[{\citenamefont{Fishman et~al.}(2022)\citenamefont{Fishman, White, and
  Stoudenmire}}]{10.21468/SciPostPhysCodeb.4}
\bibinfo{author}{\bibfnamefont{M.}~\bibnamefont{Fishman}},
  \bibinfo{author}{\bibfnamefont{S.~R.} \bibnamefont{White}}, \bibnamefont{and}
  \bibinfo{author}{\bibfnamefont{E.~M.} \bibnamefont{Stoudenmire}},
  \bibinfo{journal}{SciPost Phys. Codebases} p.~\bibinfo{pages}{4}
  (\bibinfo{year}{2022}),
  \urlprefix\url{https://scipost.org/10.21468/SciPostPhysCodeb.4}.

\bibitem[{\citenamefont{Luo et~al.}(2023)\citenamefont{Luo, Hu, Wang, W\'u, and
  Yao}}]{Yao}
\bibinfo{author}{\bibfnamefont{Z.}~\bibnamefont{Luo}},
  \bibinfo{author}{\bibfnamefont{X.}~\bibnamefont{Hu}},
  \bibinfo{author}{\bibfnamefont{M.}~\bibnamefont{Wang}},
  \bibinfo{author}{\bibfnamefont{W.}~\bibnamefont{W\'u}}, \bibnamefont{and}
  \bibinfo{author}{\bibfnamefont{D.-X.} \bibnamefont{Yao}},
  \bibinfo{journal}{Phys. Rev. Lett.} \textbf{\bibinfo{volume}{131}},
  \bibinfo{pages}{126001} (\bibinfo{year}{2023}),
  \urlprefix\url{https://link.aps.org/doi/10.1103/PhysRevLett.131.126001}.

\bibitem[{\citenamefont{Yang et~al.}(2023)\citenamefont{Yang, Wang, and
  Wang}}]{QiangHuaWang}
\bibinfo{author}{\bibfnamefont{Q.-G.} \bibnamefont{Yang}},
  \bibinfo{author}{\bibfnamefont{D.}~\bibnamefont{Wang}}, \bibnamefont{and}
  \bibinfo{author}{\bibfnamefont{Q.-H.} \bibnamefont{Wang}},
  \bibinfo{journal}{Phys. Rev. B} \textbf{\bibinfo{volume}{108}},
  \bibinfo{pages}{L140505} (\bibinfo{year}{2023}),
  \urlprefix\url{https://link.aps.org/doi/10.1103/PhysRevB.108.L140505}.

\bibitem[{\citenamefont{Zhang et~al.}(2023)\citenamefont{Zhang, Lin, Moreo, and
  Dagotto}}]{Dagotto}
\bibinfo{author}{\bibfnamefont{Y.}~\bibnamefont{Zhang}},
  \bibinfo{author}{\bibfnamefont{L.-F.} \bibnamefont{Lin}},
  \bibinfo{author}{\bibfnamefont{A.}~\bibnamefont{Moreo}}, \bibnamefont{and}
  \bibinfo{author}{\bibfnamefont{E.}~\bibnamefont{Dagotto}}
  (\bibinfo{year}{2023}), \eprint{2306.03231}.

\bibitem[{\citenamefont{Lechermann et~al.}(2023)\citenamefont{Lechermann,
  Gondolf, Botzel, and Eremin}}]{Lechermann}
\bibinfo{author}{\bibfnamefont{F.}~\bibnamefont{Lechermann}},
  \bibinfo{author}{\bibfnamefont{J.}~\bibnamefont{Gondolf}},
  \bibinfo{author}{\bibfnamefont{S.}~\bibnamefont{Botzel}}, \bibnamefont{and}
  \bibinfo{author}{\bibfnamefont{I.~M.} \bibnamefont{Eremin}}
  (\bibinfo{year}{2023}), \eprint{2306.05121}.

\bibitem[{\citenamefont{Sakakibara et~al.}(2023)\citenamefont{Sakakibara,
  Kitamine, Ochi, and Kuroki}}]{Kuroki}
\bibinfo{author}{\bibfnamefont{H.}~\bibnamefont{Sakakibara}},
  \bibinfo{author}{\bibfnamefont{N.}~\bibnamefont{Kitamine}},
  \bibinfo{author}{\bibfnamefont{M.}~\bibnamefont{Ochi}}, \bibnamefont{and}
  \bibinfo{author}{\bibfnamefont{K.}~\bibnamefont{Kuroki}}
  (\bibinfo{year}{2023}), \eprint{2306.06039}.

\bibitem[{\citenamefont{Gu et~al.}(2023)\citenamefont{Gu, Le, Yang, Wu, and
  Hu}}]{JiangPingHu}
\bibinfo{author}{\bibfnamefont{Y.}~\bibnamefont{Gu}},
  \bibinfo{author}{\bibfnamefont{C.}~\bibnamefont{Le}},
  \bibinfo{author}{\bibfnamefont{Z.}~\bibnamefont{Yang}},
  \bibinfo{author}{\bibfnamefont{X.}~\bibnamefont{Wu}}, \bibnamefont{and}
  \bibinfo{author}{\bibfnamefont{J.}~\bibnamefont{Hu}} (\bibinfo{year}{2023}),
  \eprint{2306.07275}.

\end{thebibliography}

\end{document}